\documentclass[apjl]{emulateapj}

\newcommand{\ctbd}[1]{}

\newcommand{\lc}{light curve}
\newcommand{\lcs}{light curves}
\newcommand{\Lc}{Light curve}

\newcommand{\kms}{\ensuremath{\rm km\,s^{-1}}}
\newcommand{\ms}{\ensuremath{\rm m\,s^{-1}}}

\newcommand{\gcmc}{\ensuremath{\rm g\,cm^{-3}}}
\newcommand{\ergscm}{\ensuremath{\rm erg\,s^{-1}\,cm^{-2}}}

\newcommand{\vsini}{\ensuremath{v \sin{i}}}
\newcommand{\feh}{\mathrm{[Fe/H]}}

\newcommand{\rsun}{\ensuremath{R_\sun}}
\newcommand{\msun}{\ensuremath{M_\sun}}

\newcommand{\rstar}{\ensuremath{R_\star}}

\newcommand{\teffstar}{\ensuremath{T_{\rm eff}}}

\newcommand{\loggstar}{\ensuremath{\log{g_\star}}}

\newcommand{\mearth}{\ensuremath{M_\earth}}

\newcommand{\rjup}{\ensuremath{R_{\rm J}}}
\newcommand{\mjup}{\ensuremath{M_{\rm J}}}

\newcommand{\figr}[1]{Fig.~\ref{fig:#1}}
\newcommand{\secr}[1]{\mbox{\S\ \ref{sec:#1}}}

\newcommand{\tabr}[1]{\mbox{Table~\ref{tab:#1}}}

\newcommand{\refsec}[1]{\mbox{\S\ \ref{sec:#1}}}

\newcommand{\flwof}{\mbox{FLWO 1.2 m}}

\newcommand{\flwos}{\mbox{FLWO 1.5 m}}

\newcommand{\band}[1]{\ensuremath{#1}~band}

\newcommand{\hatcurCCmag}{12.17}                                       % apparent visual magnitude
\newcommand{\hatcurCCtwomass}{2MASS~07465196+3905404}                  % 2MASS identifier
\newcommand{\hatcurCCgsc}{GSC~02959-00729}                             % GSC(1.2) identifier
\newcommand{\hatcurLCdip}{\ensuremath{12}}                             % BLS detected dip (mmag)
\newcommand{\hatcurLCrprstar}{\ensuremath{0.1050\pm0.0009}}            % Rp/R*
\newcommand{\hatcurLCimp}{\ensuremath{0.562_{-0.052}^{+0.033}}}        % impact parameter
\newcommand{\hatcurLCdur}{\ensuremath{0.1307\pm0.0013}}                % transit duration (days)
\newcommand{\hatcurLCingdur}{\ensuremath{0.0175\pm0.0013}}             % ingress/egress duration (days)
\newcommand{\hatcurLCP}{\ensuremath{4.187757\pm0.000011}}	       % period (days)
\newcommand{\hatcurLCPshort}{\ensuremath{4.1878}}                      % period (days)
\newcommand{\hatcurLCT}{\ensuremath{2454552.67168\pm0.00029}}          % epoch (BJD)
\newcommand{\hatcurSMEteff}{\ensuremath{5370\pm70}}                    % stellar effective temperature
\newcommand{\hatcurSMEzfeh}{\ensuremath{+0.05\pm0.06}}                  % stellar metallicity
\newcommand{\hatcurSMEvsin}{\ensuremath{0.7\pm0.5}}                    % stellar rotational velocity
\newcommand{\hatcurYYm}{\ensuremath{0.88\pm0.03}}                      % stellar mass
\newcommand{\hatcurYYr}{\ensuremath{1.08\pm0.04}}                      % stellar radius
\newcommand{\hatcurYYlogg}{\ensuremath{4.31\pm0.03}}                   % stellar surface gravity (refined)
\newcommand{\hatcurYYlum}{\ensuremath{0.88\pm0.09}}                    % stellar luminosity
\newcommand{\hatcurYYmv}{\ensuremath{5.06\pm0.12}}                     % stellar absolute magnitude
\newcommand{\hatcurYYvi}{\ensuremath{0.815\pm0.016}}                   % stellar V-I index
\newcommand{\hatcurYYage}{\ensuremath{14.8\pm2.0}}                     % stellar age
\newcommand{\hatcurRVK}{\ensuremath{144.9\pm2.0}}                      % RV semi-amplitude
\newcommand{\hatcurRVecosomega}{\ensuremath{+0.008\pm0.010}}           % k=e*cos(omega)
\newcommand{\hatcurRVesinomega}{\ensuremath{+0.010\pm0.013}}           % h=e*sin(omega)
\newcommand{\hatcurRVgamma}{\ensuremath{0.8\pm0.1}}                    % RV mean zero velocity
\newcommand{\hatcurPPi}{\ensuremath{86.7\pm0.4}}                       % orbital inclination
\newcommand{\hatcurPPlogg}{\ensuremath{3.33\pm0.04}}                   % planetary surface gravity (log cgs)
\newcommand{\hatcurPPar}{\ensuremath{9.67\pm0.35}}                     % relative orbital radius (a/R*)
\newcommand{\hatcurPParel}{\ensuremath{0.0488\pm0.0006}}               % semimajor axis (AU)
\newcommand{\hatcurPPrho}{\ensuremath{0.96_{-0.11}^{+0.14}}}           % planetary density (cgs)
\newcommand{\hatcurPPmlong}{\ensuremath{1.059\pm0.028}}                % planetary mass (M_jup)
\newcommand{\hatcurPPrlong}{\ensuremath{1.109\pm0.050}}                % planetary radius (R_jup)
\newcommand{\hatcurPPmrcorr}{\ensuremath{0.23}}                        % mass/radius correlation
\newcommand{\hatcurPPteff}{\ensuremath{1221\pm27}}                     % planetary temperature (K)
\newcommand{\hatcurPPtheta}{\ensuremath{0.105\pm0.005}}                % Safranov number
\newcommand{\hatcurXdist}{\ensuremath{260\pm12}}                       % distance (pc)

\newcommand{\hatcurCCtassvi}{\ensuremath{0.82\pm0.09}}  		% 
\newcommand{\hatcur}{XO-5}
\newcommand{\hatcurb}{XO-5b}
\newcommand{\hatcurCCra}{\ensuremath{7^{\mathrm{h}}46^{\mathrm{m}}51^{\mathrm{s}}.97}}%
\newcommand{\hatcurCCdec}{\ensuremath{+39^{\circ}05'40''.5}}

\shortauthors{P\'al et al.}
\shorttitle{Independent confirmation of XO-5b}

\begin{document}

%% Titlepage
\title{Independent confirmation and refined parameters of the hot Jupiter
\hatcur\lowercase{b}${}^{1}$}

%% Authors
\author{
	A.~P\'al\altaffilmark{2,3,4},
	G.~\'A.~Bakos\altaffilmark{2,5},
	J.~Fernandez\altaffilmark{2},
	B.~Sip\H{o}cz\altaffilmark{3,2},
	G.~Torres\altaffilmark{2},
	D.~W.~Latham\altaffilmark{2},
	G\'eza~Kov\'acs\altaffilmark{4},
	R.~W.~Noyes\altaffilmark{2},
	G.~W.~Marcy\altaffilmark{6},
	D.~A.~Fischer\altaffilmark{7},
	R.~P.~Butler\altaffilmark{8},
	D.~D.~Sasselov\altaffilmark{2},
	G.~A.~Esquerdo\altaffilmark{2},
	A.~Shporer\altaffilmark{9},
	T.~Mazeh\altaffilmark{9},
	R.~P.~Stefanik\altaffilmark{2},
	H.~Isaacson\altaffilmark{7}
}
\altaffiltext{1}{%
	Based in part on observations obtained at the W.~M.~Keck
	Observatory, which is operated by the University of California and
	the California Institute of Technology. Keck time has been
	granted by NOAO and NASA (programs N162Hr, N128Hr and A264Hr).
}
\altaffiltext{2}{Harvard-Smithsonian Center for Astrophysics,
	Cambridge, MA, apal@szofi.net}

\altaffiltext{3}{Department of Astronomy,
	E\"otv\"os Lor\'and University, Budapest, Hungary.}

\altaffiltext{4}{Konkoly Observatory, Budapest, Hungary}

\altaffiltext{5}{NSF Fellow}

\altaffiltext{6}{Department of Astronomy, University of California,
	Berkeley, CA}

\altaffiltext{7}{Department of Physics and Astronomy, San Francisco
	State University, San Francisco, CA}

\altaffiltext{8}{Department of Terrestrial Magnetism, Carnegie
	Institute of Washington, DC}

\altaffiltext{9}{Wise Observatory, Tel Aviv University, 
	Tel Aviv, Israel 69978}

%% EOF authors

%% abstract
\begin{abstract} 

We present HATNet observations of \hatcurb, confirming its planetary
nature based on evidence beyond that described in the announcement of
\citet{burke08}, namely, the lack of significant correlation between
spectral bisector variations and orbital phase.  In addition, using
extensive spectroscopic measurements spanning multiple seasons, we
investigate the relatively large scatter in the spectral line
bisectors. We also examine possible blended stellar configurations
(hierarchical triples, chance alignments) that can mimic the planet
signals, and we are able to show that none are consistent with the sum
of all the data.
The analysis of the $S$ activity index shows no significant stellar
activity. Our results for the planet parameters are consistent with
values in \citet{burke08}, and we refine both the stellar and planetary
parameters using our data. \hatcurb{} orbits a slightly evolved, late G
type star with mass $M_\star=\hatcurYYm$\,\msun, radius
$R_\star=\hatcurYYr$\,\rsun, and metallicity close to solar. The
planetary mass and radius are $\hatcurPPmlong\,\mjup$ and
$\hatcurPPrlong\,\rjup$, respectively, corresponding to a mean density
of $\hatcurPPrho\,\gcmc$. The ephemeris for the orbit is
$P=\hatcurLCP\,{\rm d}$, $E=\hatcurLCT$\,(BJD) with transit duration of
$\hatcurLCdur\,{\rm d}$. By measuring four individual transit centers,
we found no signs for transit timing variations. The planet \hatcurb{}
is notable for its anomalously high Safronov number, and has a high
surface gravity when compared to other transiting exoplanets with
similar period.

\end{abstract}

%% EOF abstract

%% keywords


%% EOF keywords

%% EOF titlepage

% ##########################################################################
%% Introduction
\section{Introduction}
\label{sec:intro}

There are numerous dedicated transit searches surveying the sky for
extrasolar planets that periodically transit across the face of their
host star. Among the wide angle searches, those presenting discoveries
have been TrES \citep{brown:00,dunham:04}, XO
\citep{McCullough:2005,Burke:07}, HATNet \citep{Bakos:02,Bakos:04}, and
SuperWASP \citep{pollacco2006,Cameron:2007}. The initial high hope of
finding hundreds of such planets \citep{Horne:2001} was followed by 5
years of poor harvest, and a steep learning curve for these, and many
other projects. In retrospect we now understand that several important
factors had initially been underestimated, such as the need for
dedicated telescope time, optimal precision, stable instrumentation,
low systematic noise, the number of false positives \citep{Brown:2003},
optimal follow-up strategy, and access to high precision spectroscopic
instruments. The last year showed an exponential rise in
announcements\footnote{http://www.oklo.org, http://www.exoplanet.eu},
indicating that these dedicated efforts have started to bear fruit. In
fact, they have reached a success rate such that the same object is
occasionally independently found and announced by different groups
(WASP-11b: \citet{west2008} = HAT-P-10b: \citet{bakos2008}). Such
scenarios are not necessarily duplication of effort. It is reassuring
that completely independent discoveries, follow-up observations and
analyses lead to similar parameters. They also provide an opportunity
for joint analysis of all datasets. Here we report on a similar case,
the confirmation of the planetary nature of the transiting object
\hatcurb{}, announced by \cite{burke08}. The present paper provides not
only strong new evidence supporting the planetary nature of the object,
but also improved physical properties that aid in the comparison with
theories of planet structure and formation. In \S~\ref{sec:det} we
describe the details of the photometric detection. The follow-up
observations, including the discussion of the bisector span
measurements are presented in \S~\ref{sec:fu}. The subsequent steps of
the analysis in order to characterize the star, orbit and the planet
are discussed in \S~\ref{sec:anal}.

%% EOF introduction

% ##########################################################################
%% Photometric detection
\section{Photometric detection}
\label{sec:det}

Two telescopes of the HATNet project, namely \mbox{HAT-6}, stationed at
Fred Lawrence Whipple Observatory (FLWO, $\lambda=111\arcdeg$W), and
\mbox{HAT-9}, located on the rooftop of the Submillimeter Array control
building at Mauna Kea, Hawaii ($\lambda=155\arcdeg$W), were used to
observe HATNet field ``G176'' ($\alpha = 07^{\rm h} 28^{\rm m}$,
$\delta = +37\arcdeg 30\arcmin$) on a nightly basis between 2004
November 26 and 2005 May 9. Altogether we acquired with these
telescopes 2640 and 4280 frames, respectively, with exposures of 5
minutes.

A number of candidates have emerged from this field, and have been
subjected to intense follow-up by larger instruments (\secr{fu}). One
candidate has become the transiting planet we call HAT-P-9b
\citep{shporer2008}. Another candidate internally labeled HTR176-002
has received extensive follow-up over the past two years. However, the
large scatter in the spectral line bisectors, and their tentative
correlation with orbital phase discouraged us from early announcement,
and motivated us to pursue it further. Subsequently, HTR176-002 was
announced as XO-5b by the XO group in 2007 May \citep[][hereafter
B08]{burke08}. Nevertheless, we present here our results since they
provide independent confirmation and also refine most of the
parameters.

By chance, \hatcur{} happens to fall at the edge of field ``G176''
which overlaps with field ``G177'' ($\alpha = 08^{\rm h} 00^{\rm m}$,
$\delta = +37\arcdeg 30\arcmin$). This field has been observed by the
HATNet telescope \mbox{HAT-7} and by the WHAT telescope at Wise
Observatory, Israel \citep{shporer2006}. Using these telescopes we
collected 5440 and 1930 frames, respectively. Altogether we obtained
$\sim14290$ frames with photometric information on \hatcur{} --- an
unusually rich dataset compared to data available for a typical HATNet
transit candidate.

\begin{figure}[!ht]
\plotone{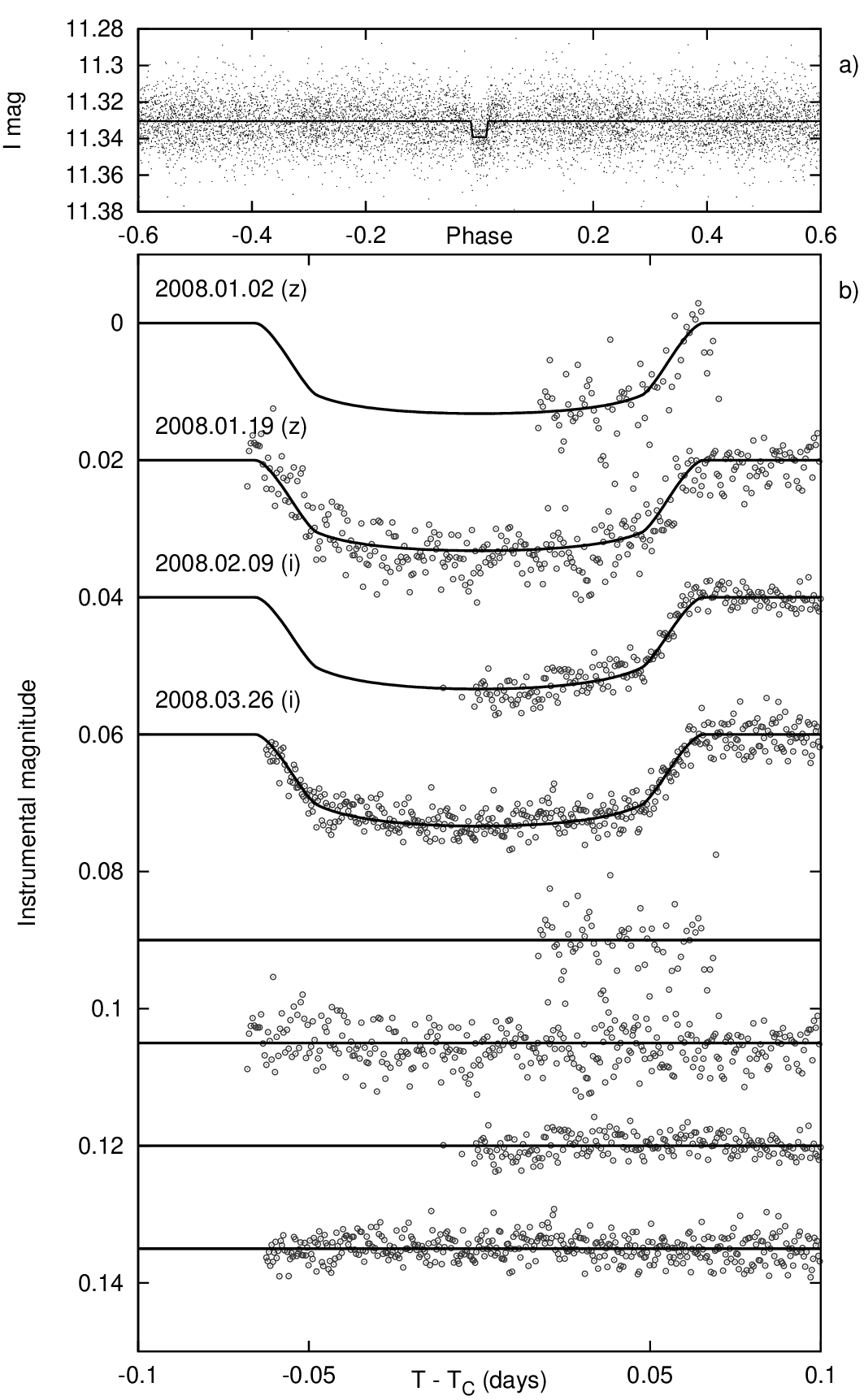}
\caption{
	({\bf a:}) The \lc{} of \hatcur{} with all 14290 points taken in
	the \band{I}, by the telescopes \mbox{HAT-6}, \mbox{HAT-7},
	\mbox{HAT-9} and WHAT. The \lc{} is folded with the period of $P =
	\hatcurLCP$\,days (which is the result of the fit described in
	\secr{anal}). The superimposed curve shows the best fit model,
	neglecting limb darkening.
	({\bf b:}) Unbinned instrumental Sloan \band{z} and \band{i}
	follow-up transit photometry light curves acquired with KeplerCam
	on the \flwof{} telescope on 2008
		January 2  ($N_{\rm tr}=0$, \band{z}), 
		January 19 ($N_{\rm tr}=4$, \band{z}), 
		February 9 ($N_{\rm tr}=9$, \band{i}) and 
		March 26   ($N_{\rm tr}=20$, \band{i}). 
	Superimposed are our best-fit transit models
	(\secr{anal}).
\label{fig:lc}}
\end{figure}

The frames from field ``G176'' were processed and analyzed as described
e.g.~in \citet{Bakos:07}. The \lcs{} from this field were corrected for
trends using the method of External Parameter Decorrelation \citep[EPD,
see][]{bakos:2009}, and the Trend Filtering Algorithm
\citep[TFA;][]{Kovacs:05}. The \lcs{} were then searched for periodic
box-like signals using the Box Least Squares algorithm of
\cite{Kovacs:02}. We detected a significant dip in the \lc{} of the
$I\approx\hatcurCCmag$ magnitude star \hatcurCCgsc{} (also known as
\hatcurCCtwomass{}; $\alpha = \hatcurCCra$, $\delta = \hatcurCCdec$;
J2000), with a depth of $\sim\hatcurLCdip$\,mmag. The period of the
signal was $P=\hatcurLCPshort$\,days, while the relative duration
(first to last contact) of the transit events was $q\approx0.027$,
which is equivalent to a total duration of $Pq\approx2.6$~hours (see
\figr{lc}a).

%% EOF Photometric detection

% ++++++++++++++++++++++++++++++++++++++++++++++++++++++++++++++++++++

% ##########################################################################
%% Follow-up observations
\section{Follow-up observations}
\label{sec:fu}

% =====================================================================
\subsection{Reconnaissance Spectroscopy}
\label{sec:rec}

In order to exclude the possibility of a false planetary detection, due
to the misinterpretation of a transit-like signal caused by another
astrophysical scenario (such as an F\,+\,M dwarf system), we observed
the candidate HTR176-002 with the CfA Digital Speedometer
\citep{Latham:92} on the \flwos{} Tillinghast reflector. We acquired
four spectra between 2007 January and March, each with an individual
precision of $0.5\,\kms$. The observations showed a mean radial
velocity of $\gamma = -10.6\,\kms$ with an rms of $0.3\,\kms$,
therefore ruling out a low-mass stellar companion (but not a triple
system), which would cause significantly higher RV variations. The
spectroscopy also yielded an estimate for the projected rotational
velocity and surface gravity of the star.

% =====================================================================
\subsection{High S/N Spectroscopy and Subsequent Analysis}
\label{sec:spec}

We obtained high resolution and high signal-to-noise spectra with the
Keck-I telescope and HIRES instrument \citep{Vogt:94}. We acquired 17
exposures with the iodine cell, and an additional iodine-free
``template''. The measurements were made between 2007 March 27 and 2008
May 17. The purpose of these observations was threefold: i) to obtain
high precision radial velocity (RV) measurements, ii) to characterize
the stellar properties, and iii) to check for spectral line bisector
variations as an indication of blends. These steps are discussed in the
following paragraphs.

\begin{figure} 
\plotone{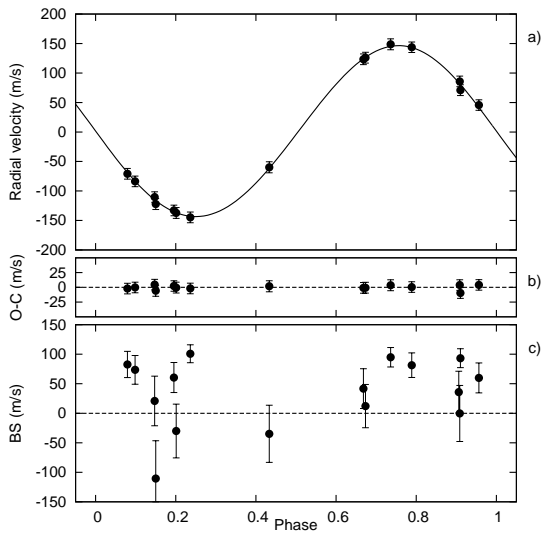}
\caption{
	(a) Radial-velocity measurements from Keck for \hatcur{}, along
	with our orbital fit (see \secr{anal}). The
	center-of-mass velocity $\gamma$ and the correlation correction
	for the bisector span variations has been subtracted.
	(b) Phased residuals after subtracting the orbital fit (also see
	\secr{anal}). The rms variation of the residuals is
	about $4.6$\,\ms.
	(c) Bisector spans (BS) for 16 of the Keck spectra (including
	the iodine-free template). 
	Note that the scales of the panels are the same.
\label{fig:rv}}
\end{figure}

As regards measuring the RV variations, the superimposed dense forest
of $\mathrm{I}_2$ absorption lines enables us to obtain an accurate
wavelength shift compared to the template observation
\citep{Marcy:92,Butler:96}. The final RV measurements and their errors
are listed in Table~\ref{tab:rvs}. The folded data, with our best fit
(see \secr{anal}) superimposed, are plotted in \figr{rv}, upper panel.

The stellar atmosphere parameters were determined using the iodine-free
template spectrum. The spectral modeling was performed using the SME
software \citep{Valenti:96}, with wavelength ranges and atomic line
data as described by \citet{Valenti:05}. We obtained the following {\em
initial} values: effective temperature $\teffstar = 5505\pm70$\,K,
surface gravity $\loggstar = 4.61\pm0.10$ (cgs), iron abundance
$\feh=+0.16\pm0.06$, and projected rotational velocity
$\vsini=0.7\pm0.5$\,\kms.

% =====================================================================
\subsection{Photometric follow-up observations}
\label{sec:photfol}

We obtained follow-up photometric observations on four nights using the
KeplerCam CCD on the \flwof{} telescope through Sloan $z$ and $i$
bands.  The observations were performed on 2008 January 2 (partial
transit), January 19 (full transit), February 9 (partial) and March 26
(full), with the total number of object frames being 114, 428, 268 and
521, respectively. The integration times used at these nights were 45,
30, 30 and 15 seconds, respectively while the readout and storage
required an additional $\sim 12$ seconds per frame.  The typical rms of
the follow-up \lcs{} was 2\,mmag at the above cadence.

We performed aperture photometry on the calibrated frames, using an
aperture series that ensures optimal flux extraction.  Details on the
astrometry, photometry, decorrelation for trends, etc., have been
discussed in, e.g., \cite{Bakos:07}. The light curves are plotted in
the lower panel of \figr{lc}, superimposed with the best-fit transit
\lc{} model (see \secr{anal}).

%% EOF Follow-up observations

% ##########################################################################
%% Blend modeling
\section{Blend analysis}
\label{sec:blend}

A stellar eclipsing binary that is unresolved from a bright source
would manifest itself as a blended system with shallow photometric
transits, and with RV variations that are of the same order of
magnitude as one can expect from a planetary system
\citep[e.g.][]{Queloz:01}. We investigated whether such a blend is a
feasible physical model for HTR176-002 in two ways: by examining the
spectral line bisectors, and with a detailed modeling of the light
curve under various possible blend scenarios.

\begin{figure} 
\plotone{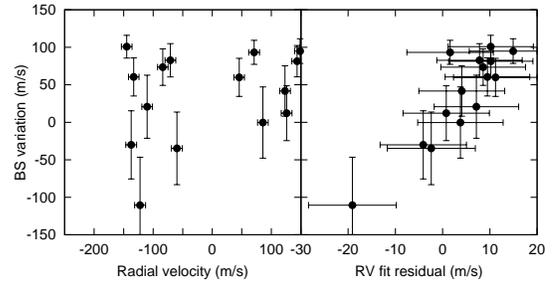}
\caption{
	Bisector span variations as a function of the RV (left panel) and
	RV fit residual (right panel). The right panel shows the fit
	residuals when the correlation term was not included in the fit.
	Note that on the graphs the horizontal scales are not the same.
\label{fig:rvbis}}
\end{figure}

For a blended eclipsing binary, in addition to the decrease in the
observed RV amplitude, the spectral lines would be distorted, as
quantified by the ``bisector spans''
\citep[see][]{Torres:05,Torres:07}. If the bisector span variations
correlate with the orbital phase, or the magnitude of these variations
is comparable with the RV amplitude, then the system is likely to be a
false positive (hierarchical triple or chance alignment with a
background binary) rather than a single star with a planetary
companion. In order to rule this out we derived the bisector spans by
cross-correlating the iodine-free ranges of the obtained spectra
against a synthetic template spectrum.  We found that the standard
deviation of the bisector spans is approximately $\sim60\,\ms$, which
is comparable to the magnitude of the RV variation itself
($K=\hatcurRVK$\,\ms; see Table~\ref{tab:parameters}).  The large
bisector variations discouraged us from publication even after the
first full transit follow-up \lc{} was obtained in January 2008, and we
continued acquiring high resolution spectroscopy to establish whether
there is any significant correlation between the bisectors and the
orbital phase (or equivalently, with the actual RV values). In
\figr{rvbis} we display our measurements of the bisector spans as the
function of both the RV and the RV residuals from the best
fit\footnote{As we will discuss later, our finally accepted best fit
values were derived by including a decorrelation factor against this
bisector span correlation. In the plot the RV residuals are shown
before subtracting this correlation term.}. There is no statistically
significant correlation between the velocities and bisector variations,
as would be expected for a blend.  However, there is apparently a
correlation between the RV \emph{residuals} and the bisector spans.
This could be due to activity on the star (e.g., spottedness), where
the activity (if periodic) causes both RV and bisector variations, but
in a way that is not commensurate with the orbital period of the
companion. We exploit this correlation in the joint analysis of the RV
and photometric data (see \S~\ref{sec:anal}) where we show that the
unbiased residual of the RV signal can be significantly decreased with
the inclusion of an additional term proportional to the bisector spans.

In order to rule out or confirm the importance of the stellar activity,
we computed the Ca II emission index $S$ \citep{Noyes:84}. The derived
indices are also shown in Table~\ref{tab:rvs}. We found that the mean
value of $S=0.16\pm0.02$ is moderately low, and the correlations
between the values of $S$ and the radial velocity data or RV fit
residuals are negligible (see also \refsec{lcrvanal}).

As a further way of assessing the true nature of the candidate, we
investigated possible blend configurations by performing light curve
fits of our highest-quality follow-up photometry (data in the Sloan
\band{i}) following the procedures described by \cite{Torres:04}.
Briefly, we attempted to reproduce the observed photometric variations
with a model based on the EBOP binary-fitting program \citep{Etzel:81,
Popper:81} in which three stars contribute light, two of which form an
eclipsing binary with the orbital period found for \hatcur. The light
from the third star (the candidate) then dilutes the otherwise deep
eclipses of the binary, reducing them to the level observed for
HTR176-002 ($\sim$1.2\% depth). The properties of the main star were
adopted from the results of our analysis below, and those of the binary
components (mass, size, brightness) were constrained to satisfy
representative model isochrones. We explored all possible combinations
for the binary components, and determined the best fits to the light
curve in a chi-square sense.

\begin{figure} 
\plotone{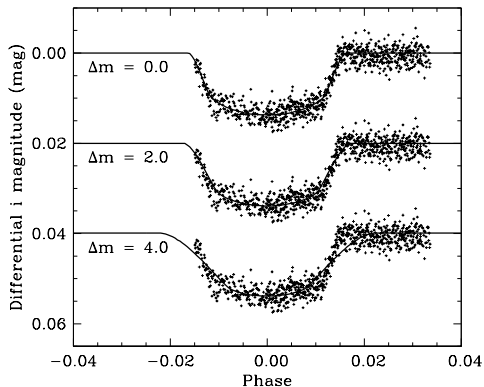}
\caption{
Blend modeling for \hatcur, based on our Sloan $i$-band
photometry. As examples we show the best fits corresponding to three
different blend scenarios, with the bottom two displaced vertically
for clarity. \emph{Top:} Model corresponding to a hierarchical triple
(see text), which is ruled out because the implied brightness of the
eclipsing binary is so large ($\sim$50\%) that our spectra would be
double-lined. \emph{Middle:} Model corresponding to a chance alignment
with a background eclipsing binary, in which the distance modulus
difference between the binary and the candidate is $\Delta m = 2$. The
ingress and egress are already seen to be too long, and the fit
implies a relative brightness of $\sim$20\% that would be easily
detectable spectroscopically. \emph{Bottom:} Chance alignment model
with $\Delta m = 4$ in which the binary is much fainter ($\sim$5\%),
but the best-fit model does not match the observations well. These
simulations rule out background blend scenarios.
\label{fig:blends}}
\end{figure}

The case of a hierarchical triple (all stars at the same distance)
yielded an excellent fit to the photometry (see top curve in
Figure~\ref{fig:blends}), but implies an eclipsing binary with a
primary that is half as bright as HTR176-002 itself. This is clearly
ruled out by our Keck spectra and even our Digital Speedometer spectra,
both of which would show obvious double lines.

We then considered scenarios in which the eclipsing binary is in the
background (which would make it fainter), and is spatially unresolved.
Because the proper motion of the candidate is relatively small
\citep[$\sim$30 mas~yr$^{-1}$;][]{Monet:03}, the chance alignment would
remain very close for decades, precluding the direct detection of the
binary in archival photographic images such as those available from the
Digital Sky Survey. For convenience we parametrized how far behind the
eclipsing binary is placed relative to the candidate in terms of the
difference in distance modulus, $\Delta m$, and we explored a wide
range of values. As an example, we find that for $\Delta m = 4$ (binary
about 1.7 kpc behind) the best fit yields a relative brightness for the
binary of only 5\%, which is at or below our detection threshold of
5--10\% from the Keck spectra. However, the ingress and egress are
clearly too long given the quality of our photometry
(Figure~\ref{fig:blends}, bottom curve). For a smaller separation of
$\Delta m = 2$ (binary some 500 pc behind) the fit is somewhat better,
though still visibly in disagreement with the observations
(Figure~\ref{fig:blends}, middle), and the relative brightness
increases to 20\%, which we would have noticed. Additional tests
changing the inclination angle from the edge-on configurations
considered above to lower angles did not alleviate the discrepancies.

The above modeling rules out both a hierarchical triple and a
background eclipsing binary as possible alternate explanations for the
photometric signals we detect. This, combined with the lack of any
clear correlation between the bisector spans and the radial
velocities, constitutes compelling evidence of the planetary nature of
HTR176-002 = XO-5, and convinces us that the scatter in the bisector
spans described above is intrinsic to the star.

%%%%%%%%%%%%%%%%%%%%%%%%%%%%%%%%%%%%%%%%%%%%%%%%%%%%%%%%%%%%%%%%%%%%%%%%%%%%%%
%%%%%%%%%%%%%%%%%%%%%%%%%%%%%%%%%%%%%%%%%%%%%%%%%%%%%%%%%%%%%%%%%%%%%%%%%%%%%%

% ##########################################################################
%% Analysis
\section{Analysis}
\label{sec:anal}

In this section we describe briefly our analysis yielding the orbital,
planetary and stellar parameters for the \hatcur{} system.

% =====================================================================
\subsection{Light curve and radial velocity analysis}
\label{sec:lcrvanal}

For the initial characterization of the spectroscopic orbit, we fitted
a Keplerian model to the Keck RV data, allowing for eccentricity by
including as adjustable parameters the Lagrangian orbital elements
$k=e\cos\varpi$ and $h=e\sin\varpi$, in addition to a velocity offset
$\gamma$, the semi-amplitude $K$ and the epoch $E$. The period $P$ was
held fixed at the value found from the HATNet \lc{} analysis (from BLS,
see above).  We found that $k$ and $h$ are insignificant compared to
their uncertainties ($k=-0.003\pm0.029$, $h=-0.009\pm0.023$),
suggesting that the orbit is circular However, in the determination of
the orbital and stellar parameters, we incorporated the uncertainties
yielded by the $k$ and $h$ orbital elements.

We proceeded next with a joint fit using all data sets, namely, the
HATNet discovery light curve, the FLWO 1.2\,m follow-up light curves,
and the Keck radial velocities along with the initial estimates of the
spectroscopic properties derived through the SME analysis.  The
follow-up light curves were modeled using the analytic formalism of
\citet{Mandel:02}, assuming quadratic limb darkening. The limb
darkening coefficients $\gamma_{1,z}$, $\gamma_{1,i}$, $\gamma_{2,z}$
and $\gamma_{2,i}$ were taken from \citet{Claret:04}, interpolating to
the values provided by the initial stellar atmospheric analysis
\secr{spec}. The adjusted parameters for the joint fit were
$T_{\mathrm{c},-270}$, the time of first transit center in the HATNet
campaign, $T_{\mathrm{c},20}$, the time of the transit center at the
last follow-up (on 2008 March 26), $m$, the out-of-transit magnitude of
the HATNet \lc{} in the \band{I}, the semi-amplitude of the radial
velocity $K$, the velocity offset $\gamma$, the Lagrangian orbital
elements $k$ and $h$, the fractional planetary radius $p\equiv R_{\rm
p}/R_\star$, the square of the impact parameter $b^2$, the quantity
$\zeta/R_\star=(2\pi/P)(a/R_\star)(1-b^2)^{-1/2}\sqrt{1-e^2}(1+h)^{-1}$
-- which is related to the duration of the transit\footnote{Here
duration is not the total duration between the first and last contact
but defined as the interval between the instances when the center of
the planet crosses the the limb of the stars inward and outward.} as
$T_{\rm dur}=2(\zeta/R_\star)^{-1}$, and the out-of-transit magnitudes
$m_{\rm c,0}$, $m_{\rm c,4}$, $m_{\rm c,9}$ and $m_{\rm c,20}$ for the
four follow-up \lcs{}. See \cite{pal2008} for a detailed discussion
about the advantages of this set of parameters. The initial values were
based on the BLS analysis, and our initial characterization of the
orbit. To obtain the best-fit values, we utilized the downhill simplex
algorithm \citep[see][]{press1992}. The uncertainties and the
correlations were determined using the Markov Chain Monte-Carlo method
\citep{Ford:06} which yields the \emph{a posteriori} distribution of
the adjusted values.

As mentioned in \refsec{blend}, we found that there is a significant
correlation between the RV residuals and the bisector spans.  This
suggests it might be possible to improve the RV fit by including an
additional term to account for this correlation. We therefore expanded
the model for the velocity variation to
\begin{equation}
v_i = \gamma + K\cdot {\rm RV_0}\left(\frac{2\pi(t_i-E)}{P},k,h\right)+C_{\rm BS}b_i \label{eq:rvmodel}
\end{equation}
where ${\rm RV_0}(\cdot,\cdot,\cdot)$ represents the \emph{base}
function for the radial velocity variations\footnote{This function has
three arguments: the mean longitude measured from the transit center
and the two Lagrangian orbital elements $k$ and $h$. It is easy to show
that if $k=h=0$, ${\rm RV_0}(\lambda,0,0)= -\sin(\lambda)$.} and $b_i$
is the actual bisector span variation for the $i$-th measurement.  We
found that when omitting the last term the unbiased residual is
$8.8$\,${\rm m\,s}^{-1}$, whereas its inclusion leads to decreased
residuals of $4.6$\,${\rm m\,s}^{-1}$, nearly a factor of two better.
We tested also whether the inclusion of a similar term in
Eq.~(\ref{eq:rvmodel}) proportional to the stellar activity index (with
a coefficient $C_{\rm S-index}$) provides any further improvement in
the fit, but found that it actually degrades the residuals slightly.
The final orbital and planetary parameters (and their uncertainties)
derived in this paper are based on the above discussed radial velocity
model function decorrelated against the bisector variations.

%%%%%%%%%%%%%%%%%%%%%%%%%%%%%%%%%%%%%%%%%%%%%%%%%%%%%%%%%%%%%%%%%%%%%%%%%%

\begin{deluxetable}{lrrrrr}
\tablewidth{0pc}
\tablecaption{Relative radial velocity, bisector span and stellar 
	activity ($S$)measurements of \hatcur{}\label{tab:rvs}}
\tablehead{
	\colhead{BJD} & 
	\colhead{RV} & 
	\colhead{\ensuremath{\sigma_{\rm RV}}} &
	\colhead{Bisec} & 
	\colhead{\ensuremath{\sigma_{\rm Bisec}}} &
	\colhead{$S$} \\
	\colhead{\hbox{($2,454,000+$)}} & 
	\colhead{(\ms)} & 
	\colhead{(\ms)} &
	\colhead{(\ms)} & 
	\colhead{(\ms)} &
}
\startdata
186.94763 & $ 269.14 $ & $ 3.21 $ & $   41.75 $ & $ 33.63 $ & $ 0.1620 $ \\ %12.0
187.94425 & $      - $ & $    - $ & $   35.88 $ & $ 35.33 $ & $ 0.1530 $ \\ %13.3
187.95384 & $ 226.59 $ & $ 3.08 $ & $   -0.36 $ & $ 47.48 $ & $ 0.1598 $ \\ % 9.6
188.95403 & $  33.07 $ & $ 2.79 $ & $   20.77 $ & $ 42.04 $ & $ 0.1589 $ \\ %10.4
216.76639 & $ 294.36 $ & $ 2.70 $ & $   81.43 $ & $ 20.90 $ & $ 0.1549 $ \\ %20.3
247.80697 & $   0.00 $ & $ 3.37 $ & $  -30.10 $ & $ 45.47 $ & $ 0.1876 $ \\ % 3.3
248.77938 & $  76.84 $ & $ 3.83 $ & $  -34.80 $ & $ 48.41 $ & $ 0.1855 $ \\ % 6.3
249.78531 & $ 268.26 $ & $ 3.36 $ & $   12.13 $ & $ 36.59 $ & $ 0.1548 $ \\ % 8.4
251.78153 & $   5.41 $ & $ 3.75 $ & $ -110.60 $ & $ 64.06 $ & $ 0.1459 $ \\ % 8.3
428.02826 & $   8.16 $ & $ 3.02 $ & $  100.80 $ & $ 15.23 $ & $ 0.1543 $ \\ %18.1
430.12240 & $ 301.03 $ & $ 3.71 $ & $   94.90 $ & $ 16.46 $ & $ 0.1549 $ \\ %14.5
455.97787 & $ 222.99 $ & $ 3.41 $ & $   93.23 $ & $ 16.03 $ & $ 0.1505 $ \\ %15.6
547.92199 & $ 224.82 $ & $ 7.25 $ & $      -\tablenotemark{a} $ & $     -\tablenotemark{a} $ & $ 0.1577 $ \\ % 8.6
548.81658 & $  79.96 $ & $ 3.08 $ & $   82.68 $ & $ 22.18 $ & $ 0.1595 $ \\ %21.6
548.89652 & $  65.98 $ & $ 2.84 $ & $   73.55 $ & $ 24.33 $ & $ 0.1576 $ \\ %21.7
602.74168 & $ 193.74 $ & $ 2.61 $ & $   59.89 $ & $ 25.39 $ & $ 0.1578 $ \\ %22.1
603.74268 & $  15.18 $ & $ 2.83 $ & $   60.49 $ & $ 25.44 $ & $ 0.1604 $ \\ %21.2
\enddata
\tablenotetext{a}{This spectrum turned out to be severely contaminated
by moonlight; however, the corresponding RV is unaffected.}
\end{deluxetable}
%%%%%%%%%%%%%%%%%%%%%%%%%%%%%%%%%%%%%%%%%%%%%%%%%%%%%%%%%%%%%%%%%%%%%%%%%%%%%%

%%%%%%%%%%%%%%%%%%%%%%%%%%%%%%%%%%%%%%%%%%%%%%%%%%%%%%%%%%%%%%%%%%%%%%%%%%%%%%
%%%%%%%%%%%%%%%%%%%%%%%%%%%%%%%%%%%%%%%%%%%%%%%%%%%%%%%%%%%%%%%%%%%%%%%%%%%%%%

%%%%%%%%%%%%%%%%%%%%%%%%%%%%%%%%%%%%%%%%%%%%%%%%%%%%%%%%%%%%%%%%%%%%%%%%%%%%%%
\begin{deluxetable}{lcl}
\tablewidth{0pc}
\tablecaption{Stellar parameters for \hatcur{} \label{tab:stellar}}
\tablehead{\colhead{Parameter}	& \colhead{Value} 		& \colhead{Source}}
\startdata
$\teffstar$ (K)\dotfill         &  \hatcurSMEteff               & SME\tablenotemark{a}          \\
$[\mathrm{Fe/H}]$\dotfill       &  \hatcurSMEzfeh               & SME                           \\
$v \sin i$ (\kms)\dotfill       &  \hatcurSMEvsin               & SME                           \\
$M_\star$ ($M_{\sun}$)\dotfill  &  \hatcurYYm                   & Y$^2$+LC+SME\tablenotemark{b} \\
$R_\star$ ($R_{\sun}$)\dotfill  &  \hatcurYYr                   & Y$^2$+LC+SME                  \\
$\loggstar$ (cgs)\dotfill       &  \hatcurYYlogg                & Y$^2$+LC+SME                  \\
$L_\star$ ($L_{\sun}$)\dotfill  &  \hatcurYYlum                 & Y$^2$+LC+SME                  \\
$M_V$ (mag)\dotfill             &  \hatcurYYmv                  & Y$^2$+LC+SME                  \\
Age (Gyr)\dotfill               &  \hatcurYYage                 & Y$^2$+LC+SME                  \\
Distance (pc)\dotfill           &  \hatcurXdist                 & Y$^2$+LC+SME
\enddata
\tablenotetext{a}{SME = `Spectroscopy Made Easy' package for analysis
of high-resolution spectra \cite{Valenti:96}. See text.}
\tablenotetext{b}{Y$^2$+LC+SME = Yale-Yonsei isochrones \citep{Yi:01},
\lc{} parameters, and SME results.}
\end{deluxetable}

% =====================================================================
\subsection{Stellar and planetary parameters}
\label{sec:stellarparameters}

%%%%%%%%%%%%%%%%%%%%%%%%%%%%%%%%%%%%%%%%%%%%%%%%%%%%%%%%%%%%%%%%%%%%%%%%%%%%%%
%%%%%%%%%%%%%%%%%%%%%%%%%%%%%%%%%%%%%%%%%%%%%%%%%%%%%%%%%%%%%%%%%%%%%%%%%%%%%%

\begin{figure}
\plotone{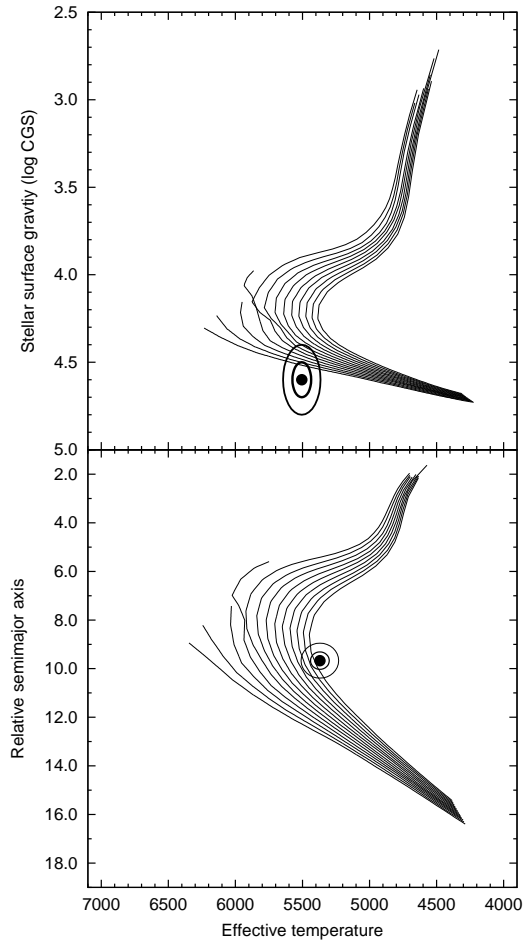}
\caption{
Stellar evolution isochrones from the Yonsei-Yale models, corresponding
to ages between $2$ and $14$\,Gyr (in steps of $1$\,Gyr), as a function
of both surface gravity (top) and normalized semimajor axis
$a/R_{\star}$ (bottom).  In the top panel the isochrone metallicity
($[\rm{Fe/H}]=+0.16$), spectroscopic surface gravity, and temperature
are from our initial SME analysis, the latter two shown with 1-$\sigma$
and 2-$\sigma$ confidence ellipsoids. In the lower panel the
metallicity ($[\rm{Fe/H}]=+0.05$), temperature, and $a/R_{\star}$ are
from the iterative analysis described in the text. Note that the latter
quantities result in a significantly different evolutionary state for
the star.
}
\label{fig:isochrones}
\end{figure}

The stellar parameters were determined in an iterative way as follows.
As pointed out by \cite{Sozzetti:07}, the stellar density is a better
luminosity indicator than the spectroscopic value of $\loggstar$. In a
first order approximation the density is related to the observable
quantities $P$ and $a/R_\star$ as
$\rho_\star=(3\pi)G^{-1}P^{-2}(a/R_\star)^3$. We used the values of
$\teffstar$ and $\feh$ from the SME analysis, together with the
distribution of $\rho_\star$ (derived from $a/R_\star$) to estimate the
stellar parameters from the Yonsei-Yale evolution models, as published
by \cite{Yi:01} and \cite{Demarque:04}. This resulted in \emph{a
posteriori} distributions of those stellar parameters, including the
mass, radius, age, luminosity and colors. From the mass and radius
distributions, we obtained a new value and uncertainty for the stellar
surface gravity: $\hatcurYYlogg$. Since this value is significantly
smaller than the previous value based on the SME analysis \secr{spec},
we repeated the atmospheric modeling by fixing the surface gravity to
the new value ($\hatcurYYlogg$), and allowing only the metallicity and
effective temperature to vary. This next iteration of the SME analysis
yielded $\teffstar=\hatcurSMEteff$\,K and $\feh=\hatcurSMEzfeh$. Based
on these new atmospheric parameters, the limb darkening coefficients
were re-calculated and we repeated the joint fit for the \lc{} and RV
parameters, followed by the stellar evolution modeling once again, in
the same way as discussed earlier. In this iteration the surface
gravity barely changed ($\loggstar=4.33\pm0.04$), so the stellar
previous parameters were accepted as final (\tabr{stellar}). In
Fig.~\ref{fig:isochrones}, we plot the evolutionary isochrones as the
function of the effective temperature and both the stellar surface
gravity and $a/R_\star$ (these are used as luminosity indicators). The
temperature, surface gravity and relative semimajor axis values
discussed here are also superimposed on these isochrone plots.

%% %% %% %% %% %% %% %% %% %% %% %% %% %% %% %% %% %% %% %% %% %% %% %% %% %% 
%% %% %% %% %% %% %% %% %% %% %% %% %% %% %% %% %% %% %% %% %% %% %% %% %% %% 

The results from this second global fit to all the available data
(photometry, radial velocities)
are listed in \tabr{parameters}. In addition, values for
some auxiliary parameters in this fit are:
$T_{\mathrm{c},-270}=2453338.22311\pm0.00236$~(BJD), 
$T_{\mathrm{c},20}  =2454552.67174\pm0.00029$~(BJD), 
$m=11.33042\pm0.00010$\,mag
and the Keck velocity offset is $\gamma=\hatcurRVgamma$\,\ms. The
best-fit values and uncertainties for the fitted parameters are
straightforward to obtain from the MC distributions. These, in turn,
lead to the planetary parameters and their uncertainties by using a
direct combination of the \emph{a posteriori} parameter distributions
of the light curve, radial velocity and stellar parameters. We find
that the mass of the planet is $M_p=\hatcurPPmlong$\,\mjup, the radius
is $R_p=\hatcurPPrlong$\,\rjup{} and its density is
$\rho_p=\hatcurPPrho$\,\gcmc. These quantities are also collected in
Table~\ref{tab:parameters}.  The correlation coefficient $C(M_p,R_p)$
between the planetary mass and radius is listed as well. We also
estimated the individual transit centers of the four follow-up \lcs{},
by adjusting only the the light curve parameters ($R_{\rm p}/R_\star$,
$b^2$, $\zeta/R_\star$, out-of-transit magnitudes) while the transit
centers were not constrained by a given epoch and period. We obtained
that the individual transit centers do not differ significantly from
the interpolated transit centers (derived from the results of the joint
fit), i.e.  the available data do not show any signs for transit timing
variations.  The independently fitted transit centers for the events
$N_{\rm tr}=0$, $N_{\rm tr}=4$ and $N_{\rm tr}=20$ differ from the
linearly interpolated values by less than $1.5$-$\sigma$, and the
difference at the event $N_{\rm tr}=9$ is nearly $2.3$-$\sigma$. The
independently fitted and the interpolated transit instants are shown in
Table~\ref{tab:tc}.

Using our best fit model, we also checked the amplitude of the
out-of-transit variations of the HATNet \lc{}, by performing a Fourier
analysis on the fit residuals. We found no significant variation in the
stellar flux, and all Fourier amplitudes were less than $0.7$\,mmag. 
This estimation gives an upper limit for the stellar activity, and is
in line with the small $S$ values derived from spectroscopy ($S\lesssim
0.186$, see Table~\ref{tab:rvs}). It is somewhat surprising that in
spite of the small activity based on the spectroscopic $S$ index, the
\lc{} out-of-transit variation, and the low \vsini{} rotational
velocity of the star, the bisector spans exhibit such a large scatter.

The Yonsei-Yale evolutionary models also provide the absolute
magnitudes and colors for different photometric bands. We compared the
$V-I$ model color with the observed TASS color
\citep[see][]{Droege:06}. Since $(V-I)_{\rm YY}=\hatcurYYvi$ and
$(V-I)_{\rm TASS}=\hatcurCCtassvi$, we conclude that the star is not
significantly affected by interstellar reddening (also note the
Galactic latitude of \hatcur{}, which is $b=26^\circ.9$). Therefore,
for the distance determination we use the distance modulus $V_{\rm
TASS}-M_V=7.18\pm0.13$, which corresponds to $d=\hatcurXdist$\,pc.

% ---------------------------------------------------------------------
\begin{deluxetable}{ll}
\tablewidth{0pc}
\tablecaption{Orbital and planetary parameters\label{tab:parameters}}
\tablehead{\colhead{~~~~~~~~~~~~Parameter~~~~~~~~~~~~} & \colhead{Value}}
\startdata
\sidehead{\Lc{} parameters}
~~~$P$ (days)                                   \dotfill        & $\hatcurLCP$          \\
~~~$E$ (${\rm BJD}$)                            \dotfill        & $\hatcurLCT$          \\
~~~$T_{14}$ (days)\tablenotemark{a}             \dotfill        & $\hatcurLCdur$        \\
~~~$T_{12} = T_{34}$ (days)\tablenotemark{a}    \dotfill        & $\hatcurLCingdur$     \\
~~~$\zeta/R_\star$ ($\mathrm{day^{-1}}$)        \dotfill        & $17.779\pm0.091$      \\
~~~$a/R_\star$                                  \dotfill        & $\hatcurPPar$         \\
~~~$R_p/R_\star$                                \dotfill        & $\hatcurLCrprstar$    \\
~~~$b \equiv a \cos i/R_\star$                  \dotfill        & $\hatcurLCimp$        \\
~~~$i$ (deg)                                    \dotfill        & $\hatcurPPi$ \phn     \\
\sidehead{Spectroscopic parameters}
~~~$K$ (\ms)                                    \dotfill        & $\hatcurRVK$          \\
%~~~$\gamma$ (\kms)                             \dotfill        & $\hatcurRVgamma$      \\
~~~$C_{\rm BS}$					\dotfill	& $0.125\pm0.025$ 	\\
~~~$C_{\rm S-index}$				\dotfill	& $0$ (adopted) 	\\
~~~$k\equiv e\,\cos\omega$                      \dotfill        & $\hatcurRVecosomega$  \\
~~~$h\equiv e\,\sin\omega$                      \dotfill        & $\hatcurRVesinomega$  \\
\sidehead{Planetary parameters}
~~~$M_p$ ($\mjup$)                              \dotfill        & $\hatcurPPmlong$      \\
~~~$R_p$ ($\rjup$)                              \dotfill        & $\hatcurPPrlong$      \\
~~~$C(M_p,R_p)$                                 \dotfill        & $\hatcurPPmrcorr$     \\
~~~$\rho_p$ (\gcmc)                             \dotfill        & $\hatcurPPrho$        \\
~~~$a$ (AU)                                     \dotfill        & $\hatcurPParel$       \\
~~~$\log g_p$ (cgs)                             \dotfill        & $\hatcurPPlogg$       \\
~~~$T_{\rm eq}$ (K)                             \dotfill        & $\hatcurPPteff$       \\
~~~$\Theta$                                     \dotfill        & $\hatcurPPtheta$      
\enddata
\tablenotetext{a}{\ensuremath{T_{14}}: total transit duration, time
between first and last contact; \ensuremath{T_{12}=T_{34}}:
ingress/egress time, time between first and second, or third and fourth
contact.}
\end{deluxetable}
% ---------------------------------------------------------------------

\begin{deluxetable}{rrr}
\tablewidth{0pc}
\tablecaption{Individual transit center measurements\label{tab:tc}}
\tablehead{
	\colhead{Event} & 
	\colhead{\ensuremath{T_{\rm C} (BJD)\tablenotemark{a}}} & 
	\colhead{\ensuremath{T_{\rm C} (BJD)\tablenotemark{b}}}\\
	\colhead{\#} & 
	\colhead{\hbox{~~~~(2,454,000$+$)~~~~}} & 
	\colhead{\hbox{~~~~(2,454,000$+$)~~~~}} 
}
\startdata
0 	& $468.91868\pm0.00181$ & $468.91666\pm0.00028$ \\
4 	& $485.66932\pm0.00058$ & $485.66768\pm0.00028$ \\
9	& $506.60475\pm0.00057$ & $506.60645\pm0.00027$ \\
20	& $552.67152\pm0.00041$ & $552.67174\pm0.00029$ \\
\enddata
\tablenotetext{a}{Derived frp, the individually fitted the transit centers
while the other light curve parameters were constrained to be equal.}
\tablenotetext{b}{Derived by interpolation from the joint fit results,
assuming a constant period.}
\end{deluxetable}

%%%%%%%%%%%%%%%%%%%%%%%%%%%%%%%%%%%%%%%%%%%%%%%%%%%%%%%%%%%%%%%%%%%%%%%%%%%%%%

%% EOF Analysis

\begin{figure*}[!ht]
\plottwo{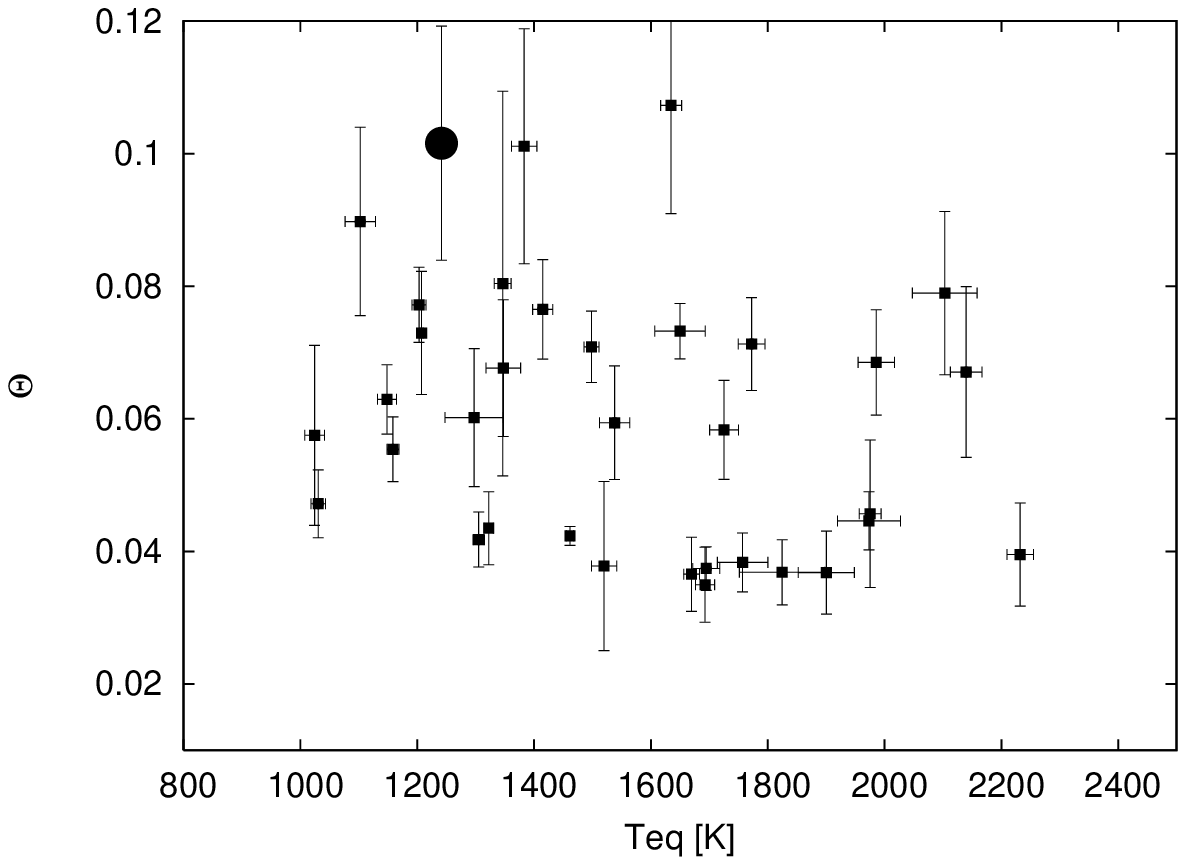}{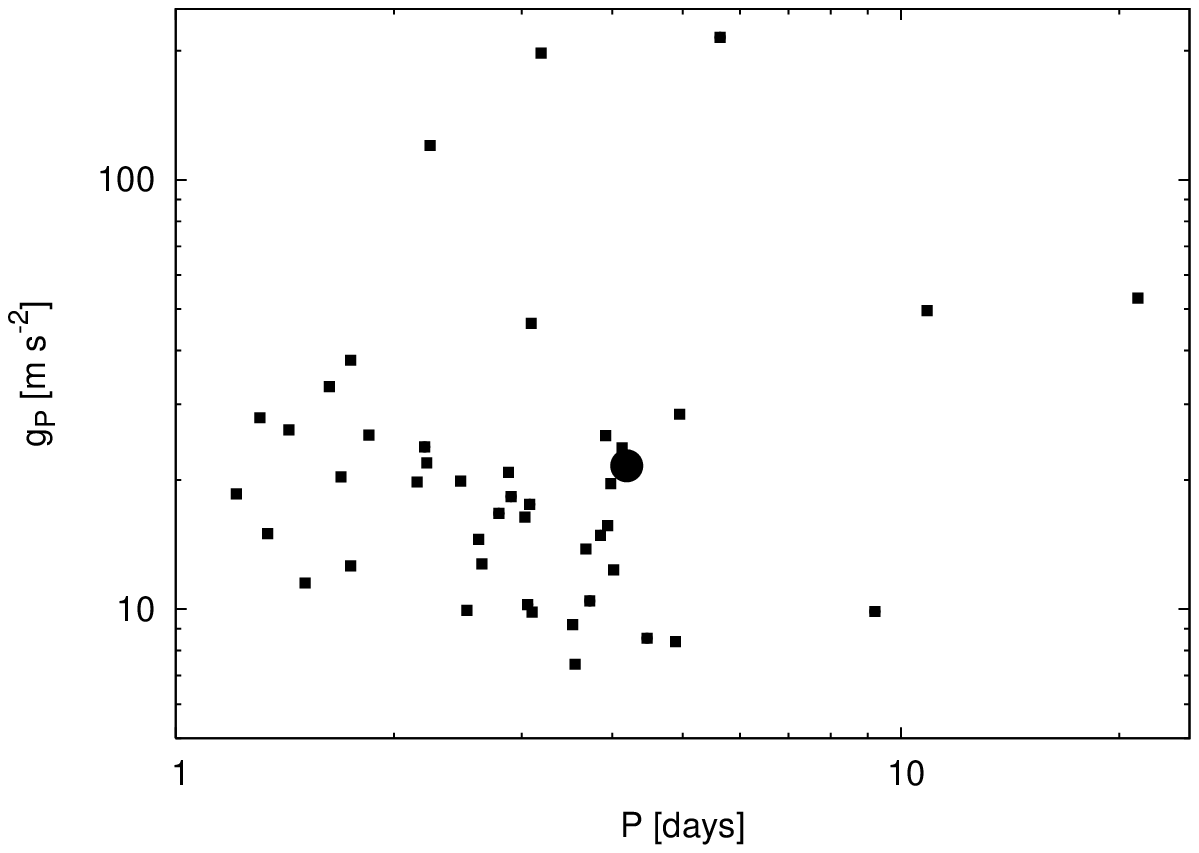}
\caption{
	(a): Safronov number vs.~equilibrium temperature for the known
	transiting extrasolar planets. \hatcurb{} is marked with a larger
	dot and it is located at the upper envelope of the Class~I
	distribution of planets (\hatcurb{} $> 0.05$). (b): Surface
	gravities as the function of the orbital period for the known
	extrasolar planets. With its relatively high surface gravity and
	orbital period, \hatcurb{} falls slightly off the main
	distribution.
\label{fig:Pg_tsaf}}
\end{figure*}

% ##########################################################################
%% Discussion
\section{Discussion}  

In this paper we have described our independent detection of the
transiting planet \hatcurb{} using the HATNet observations. A
significant component of our effort has been to examine possible
astrophysical false positives and to model the data in detail in order
to rule them out. In this way we have provided new and crucial support
for the planetary nature of the object. We also present refined values
for the system parameters. It is reassuring that the planetary
parameters in B08 and this work are consistent within 1-$\sigma$. This,
however, is somewhat coincidental, since the stellar parameters are
quite different. Based on our SME analysis, we derive a lower effective
temperature (\teffstar = \hatcurSMEteff\,K as compared to
$5510\pm44$\,K in B08), and a lower metallicity ($\feh =
\hatcurSMEzfeh$ vs.~$0.25\pm0.03$). The difference is attributed to our
iterations on the SME analysis and the transit-fitting, using the
$a/\rstar$ based mean stellar density as a luminosity indicator, and
fixing the corresponding $\loggstar$ in the SME analysis (i.e.~solving
only for $\feh$ and $\teffstar$). We derive a smaller stellar mass:
$\hatcurYYm\,\msun$ vs.~$1.0\pm0.03,\msun$, based on the same
\cite{Yi:01} isochrones. Due to the high precision photometric and RV
data, we are able to refine the planetary and orbital parameters of the
system, and decrease the uncertainties typically by a factor of
$\sim2-3$.

Based on the models of \citet{liu:08}, after re-scaling the semi-major
axis to match the insolation flux \hatcurb{} would have if it orbited a
G2V dwarf ($a_{equiv} = 0.05313$\,AU), the measured mass and radius of
\hatcurb{} require a small core to be consistent with theory even if no
internal heating is assumed. Using the work of \cite{fortney:07},
\hatcurb{} is consistent with a 300\,Myr old planet with a
50\,\mearth{} core, a 1\,Gyr old planet with a 25\,\mearth{} core, or a
4.5\,Gyr planet with a core smaller than 10\,\mearth{} mass. The
incident flux on \hatcurb{} is $\sim 4.83\cdot 10^8\ergscm$. This
corresponds to a pL class planet, based on the definitions of
\citet{fortney:08}, although it falls fairly close to the transition
area between the pL and pM classes.

We confirm that the planet has a remarkably high Safronov number,
$\Theta\equiv 1/2(V_{esc}/V_{orb})^2 = \hatcurPPtheta$, placing it at
the high end of the Class~I planets as defined by \cite{hansen2007}.
The plot of the Safronov numbers for the known TEPs as a function of
equilibrium temperature is displayed on \figr{Pg_tsaf}a. We also
confirm that \hatcurb{} has an anomalously high surface gravity, as
compared to other TEPs with similar period \citep{southworth07}.

Altogether, \hatcur{} appears to be an interesting system exhibiting a
number of anomalies including non-trivial bisector span variations, and
anomalously high Safronov number and surface gravity. Future
observations and theoretical work are required to understand these
properties.

%% EOF Discussion

% =====================================================================
%% Acknowledgements
\acknowledgements 

HATNet operations have been funded by NASA grants NNG04GN74G,
NNX08AF23G and SAO IR\&D grants. Work of G.\'A.B.~was supported by the
Postdoctoral Fellowship of the NSF Astronomy and Astrophysics Program
(AST-0702843). We acknowledge partial support also from the Kepler
Mission under NASA Cooperative Agreement NCC2-1390 (D.W.L., PI).
G.K.~thanks the Hungarian Scientific Research Foundation (OTKA) for
support through grant K-60750. A.P.~is grateful for the SAO Visiting
Student Fellowship that supported his work. This research has made use
of Keck telescope time granted through NASA and NOAO (programs N162Hr,
N128Hr and A264Hr).

%% EOF Acknowledgements

% =====================================================================
%% Bibliography

%% EOF Bibliography

\end{document}